# Field-free superconducting diode effect and magnetochiral anisotropy in FeTe$_{0.7}$Se$_{0.3}$ junctions with the inherent asymmetric barrier

*Shengyao Li[1†], Ya Deng[2†], Dianyi Hu[2], Chao Zhu[3], Zherui Yang[1], Wanghao Tian[4], Xueyan Wang[1], Ming Yue[5], Qiong Wu[5\*], Zheng Liu[2,6\*], Xiao Renshaw Wang[1,4\*]*

[1]Division of Physics and Applied Physics, School of Physical and Mathematical Sciences, Nanyang Technological University, Singapore 637371, Singapore

[2]School of Materials Science and Engineering, Nanyang Technological University, Singapore 639798, Singapore

[3]SEU-FEI Nano-Pico Center, Key Laboratory of MEMS of Ministry of Education, Collaborative Innovation Center for Micro/Nano Fabrication, Device and System, Southeast University, Nanjing 210096, China

[4]School of Electrical and Electronic Engineering, Nanyang Technological University, Singapore 639798, Singapore

[5]Faculty of Materials and Manufacturing, Key Laboratory of Advanced Functional Materials, Ministry of Education of China, Beijing University of Technology, Beijing, 100124, China

[6]CINTRA CNRS/NTU/THALES, UMI 3288, Research Techno Plaza, Singapore 637553, Singapore

[†] These authors contributed equally
\* Emails: wuqiong0506@bjut.edu.cn; z.liu@ntu.edu.sg; renshaw@ntu.edu.sg

**Abstract**

Nonreciprocal electrical transport, characterized by an asymmetric relationship between current and voltage, plays a crucial role in modern electronic industries. Recent studies have extended this phenomenon to superconductors, introducing the concept of the superconducting diode effect (SDE). The SDE is characterized by unequal critical supercurrents along opposite directions. Due to the requirement on broken inversion symmetry, the SDE is commonly accompanied by electrical magnetochiral anisotropy (eMCA) in the resistive state. Achieving a magnetic field-free SDE with field tunability is pivotal for advancements in superconductor devices. Conventionally, the field-free SDE has been achieved in Josephson junctions by intentionally intercalating an asymmetric barrier layer. Alternatively, internal magnetism was employed. Both approaches pose challenges in the selection of superconductors and fabrication processes, thereby impeding the development of SDE. Here, we present a field-free SDE in $FeTe_{0.7}Se_{0.3}$ (FTS) junction with eMCA, a phenomenon absent in FTS single nanosheets. The field-free property is associated with the presence of a gradient oxide layer on the upper surface of each FTS nanosheet, while the eMCA is linked to spin-splitting arising from the absence of inversion symmetry. Both the SDE and eMCA respond to magnetic fields with distinct temperature dependencies. This work presents a versatile and straightforward strategy for advancing superconducting electronics.

**1. Introduction**

Nonreciprocal charge transport refers to an asymmetric relationship between current and voltage. This characteristic has a long history in semiconductor pn junctions. A similar phenomenon has been extended to symmetry-breaking conductors with strong spin-orbit coupling and low Fermi level[1], where the resistance exhibits dependence on the current and magnetic field, known as electrical magnetochiral anisotropy (eMCA)[2]. Especially, this nonreciprocal response is significantly enhanced in noncentrosymmetric superconductors by several orders due to the introduction of a superconducting gap[3,4], as exemplified in superconductors with structural polarity[5,6], chirality[7], and topological surface states[8,9]. Therefore, the nonreciprocal transport in superconductors serves as an effective approach for elucidating the structural, as well as electronic properties[3]. Taking advantage of the directional resistance responses to magnetic fields and current, the effect emerges as a potential alternative to conventional semiconductor junctions for rectification applications at low temperatures[10,11].

Nonreciprocal electrical transport in noncentrosymmetric superconductors manifests as electrical magnetochiral anisotropy (eMCA) in the resistive state, where Cooper pairs start to form but coherent superconductivity is not reached. This phenomenon is attributed to the vortex motion at a lower temperature[5,12], and paraconductivity around $T_c$[13]. Recently, another manifestation of nonreciprocal transport has been identified at low temperature zone of the superconducting transition. This phenomenon is distinguished by a nonreciprocal supercurrent in the zero-resistance state, hence termed the superconducting diode effect (SDE)[14]. Characterized by an asymmetric voltage-current (*V-I*) relationship, with critical currents ($I_c$) differing for currents of opposite directions, SDE allows a fast switching between a zero-resistance and a resistive state. This feature renders SDE advantageous due to its high-rectification ratio and dissipation-less properties. Despite different origins and phenomena in SDE and eMCA, the two phenomena share a similar dependence on external magnetic fields[15]. This occurs as magnetic fields modulate the band dispersion in inversion symmetry breaking systems with spin-splitting[16,17], and through the orbital effect in chiral structures without SOC[18]. In most cases, nonreciprocal resistance from eMCA and asymmetric $I_c$ from SDE coexist, but at different temperature zones.

Studies on SDE have primarily focused on Josephson junctions[19–21] and Rashba-type superconductors[14], where the effect can be modulated by tuning Andreev-bound states and Rashba-type spin-splitting, respectively. The former is typically constructed with two superconductor electrodes bridged by a non-superconducting barrier[21], while the latter is created through the interfacial electric field in superconductor heterostructures[14]. Moving forward, the latest research highlights the field-free characteristic of the SDE, while the simultaneous magnetic field tunability enhances its applicability. Current studies of field-free SDE focus on superconductor junctions with asymmetric barriers[22] or internal magnetism[23]. However, the presence of an asymmetric barrier layer imposes meticulous requirements on the uniformity and thickness of the material[24]. Furthermore, internal magnetism presents constraints on material selection and complicates the arrangement of devices. Therefore, a straightforward approach to achieving field-free SDE with tunable nonreciprocal transport behaviour could facilitate the advancement of superconducting diode devices.

Here, we present a field-free SDE in FeTe$_{0.7}$Se$_{0.3}$ (FTS) junction accompanied with eMCA. The device is constructed utilizing chemical vapour deposition (CVD) synthesized FTS nanosheets. While the FTS single nanosheet demonstrates a typical *V-I* relationship, devoid of SDE or eMCA characteristics, the FTS junction exhibits a field-free SDE at low temperatures, along with eMCA behaviour around the critical temperature ($T_c$). The emergence of field-free SDE is attributed to the spontaneously formed gradient oxidized barrier layer on the upper surface of each FTS nanosheet. Additionally, the magnetic field and current-dependent eMCA are ascribed to the band dispersion within the rotational stacking of FTS nanosheets. These findings offer a straightforward method for achieving a field-free SDE accompanied by eMCA using thin superconductor flakes.

**2. Results and Discussion**

Iron-based superconductors, especially FeTe$_{1-x}$Se$_x$, have attracted great interest due to their high superconducting transition temperature and intriguing surface state[25–28]. As a result, it is an ideal material for constructing superconducting devices. Figure 1a schematically illustrates the fabrication process of an FTS junction. During the CVD growth process[29], a mixture of Fe$_2$O$_3$/FeCl$_2$ acts as the iron precursor, and Te, Se powders work as the chalcogen precursors. The synthesized nanosheets exhibit a uniform elemental distribution, with the atomic ratio of Te/Se measured to be 74:26 across various nanosheets (Figure S1, Supporting Information). The synthesized nanosheets on the SiO$_2$ substrate are transferred to the top of another FTS nanosheet, forming an FTS junction. The atomic-resolution scanning transmission electron microscopy (STEM) indicates a well-crystalized tetragonal phase structure of the synthesized nanosheet as shown in Figure 1b.

Electrical measurements are performed to confirm the superconductivity of the synthesized FTS nanosheet, using the FTS single nanosheet device in Figure 1c. The temperature-dependent resistance of the FTS nanosheet exhibits metallic behaviour upon cooling, with a $T_c$ of 12.6 K (Figure 1d). Figure 1g shows the optical image of the FTS junction, constructed using FTS thin nanosheets of different thicknesses (Figure 1e). Figure 1f illustrates the measurement setup of the FTS junction. Current is applied at two ends of the device, and voltages across the junction are recorded. The FTS junction demonstrates a similar metallic behaviour upon cooling, with a $T_c$ of 12 K (Figure 1h), which is slightly lower than the single nanosheet.

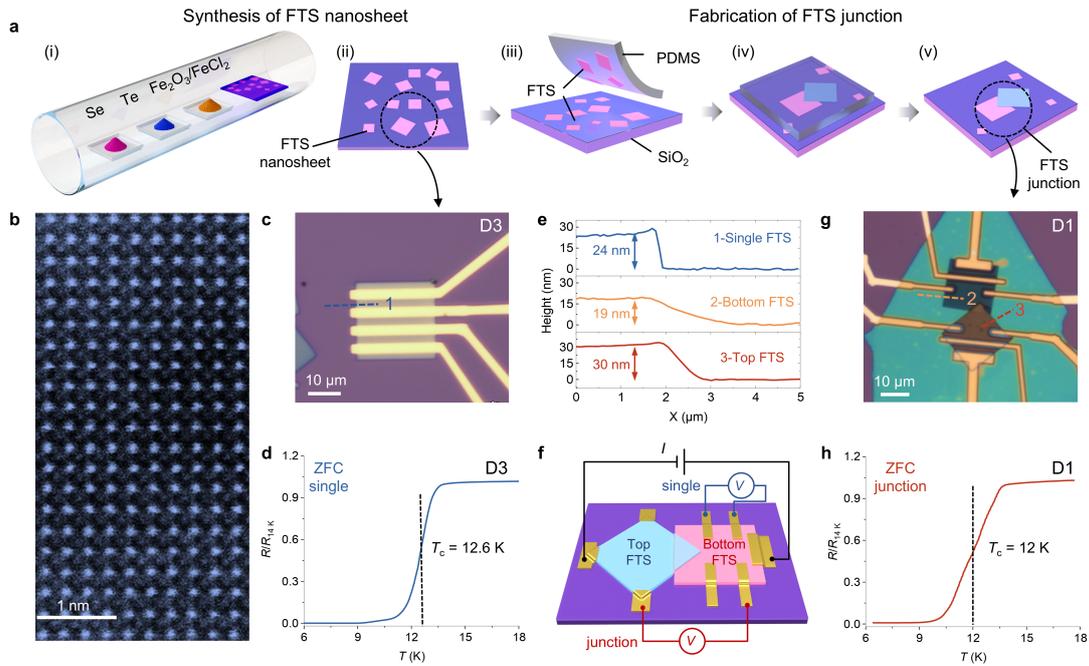

Figure 1. Fabrication and characterization of the FTS junction. (a) Schematic of constructing a junction device using the CVD synthesized FTS nanosheets. (b) Atomic-resolution STEM image of the FTS nanosheet from the *c*-axis. (c) Optical image of a single FTS nanosheet, scale bar is 10 µm. (d) The temperature-dependent relative resistance ($R/R_{14\,K}$) of the single nanosheet. (e) Atomic force microscope (AFM) height profile of the nanosheets along the dashed lines in optical images. (f) Measurement configuration of the FTS junction. (g) Optical image of the FTS junction, scale bar is 10 µm. (h) Temperature-dependent relative resistance ($R/R_{14\,K}$) of the junction.

The superconductivity-induced *V-I* property displays temperature dependence. Figures 2a-c comparatively investigate the *V-I* characteristics of the FTS junction contrasted with the FTS single nanosheet. Here, a positive sweep denotes the current sweep in the sequence of zero to positive (0–p), positive to negative (p–n), and negative back to zero (n–0). Conversely, a negative sweep refers to the current sweep in the reversed direction, following the order of zero to negative (0-n), negative to positive (n-p), and positive back to zero (p-0). The illustration of current sweeps in different directions can be found in Figure S2 of the Supporting Information.

Figure 2a shows the temperature-dependent *V-I* relationship of the FTS junction. The lower panel depicts the *V-I* relationship of the FTS junction obtained by positively sweeping the current at 5 K. The red curve denotes the current sweep from n to p, while the blue curve denotes the current sweep from p to n. A hysteretic *V-I* loop with four branches is observed, attributed to the resistively and capacitively shunted junction (RCSJ) model[30]. This model gives rise to the current that breaks superconductivity, $I_c$, and the current that returns to the superconducting state, the reversal current, $I_r$. To illustrate its temperature dependence, the upper panel of Figure 2a presents the voltage contour plot of the FTS junction in the n to 0 and 0 to p branches. The definition of four critical currents and *V-I* curves at each temperature are illustrated in Figure S3 in Supporting Information. The dashed line indicates the voltage sudden jumps in each *V-I* curve. The blue dashed lines at higher currents indicate the breaking of superconductivity of FTS single nanosheets, denoted as $I_c$ (single), which exhibit a symmetric arrangement along the current axis. The red dashed line on the left side represents the reversal current, $I_r(-)$, while the red dashed line on the right side represents the critical current, $I_c(+)$. With the increase in temperature, $I_r(-)$ and $I_c(+)$ converge with $I_c$ (single) at $T$ = 10.5 K.

Figure 2b shows the temperature-dependent *V-I* relationship of the FTS single nanosheet. The upper panel displays the voltage contour plot of the top 30 nm FTS single nanosheet. The single nanosheet demonstrates a typical symmetric *V-I* characteristic, with similar $I_c$ for currents of opposite directions. The lower panel specifies the *V-I* curve at 5 K. The red curve represents the current sweep from n to p, while the blue curve represents the current sweep from p to n. These two curves overlap with each other, with no branches or hysteresis observed in FTS single nanosheets.

Figure 2c summarizes the temperature-dependent $I_c$ of the FTS junction (red dots) and single nanosheet (blue dots). The $I_c$ of the FTS single nanosheet is nearly two orders of magnitude larger than that of the FTS junction, hence, the lower panel zooms in on the characteristic of FTS junction. Both $I_c$ values increase with decreasing temperature and gradually saturate at the lowest temperature. The $I_c$ of the FTS single nanosheet is consistent with other reported records[31], thus providing further evidence for the quality of the CVD-synthesized FTS nanosheet. The $I_c$ of the FTS junction is an important parameter for estimating the gap parameter of the superconducting junction, calculated as the product of $I_c$ and $R_N$. The $I_c R_N$ values of different FTS junction devices yield similar results (Figure S4, Supporting Information), indicating the reproducibility of the FTS junction fabrication[32,33].

To comparatively study the critical currents under opposite current biases, Figure 2d presents the *V-I* curve at $T$ = 2 K in a positive sweep without an external magnetic field. Absolute values of voltages and currents are taken for the 0 to n (blue solid line) and n to 0 (blue dashed line) branches. $I_c(+)$ is not equal to its counterpart, $I_c(-)$. The difference, $\Delta I_c$ ($\Delta I_c = |I_c(-)| - |I_c(+)|$), is around 15 μA, indicating a field-free SDE in FTS junction: For positive and negative currents with amplitudes in the purple region, the device switches between superconducting and resistive states. The asymmetric *V-I* relationship derived from the RCSJ model results in distinct resistive states under positive and negative currents, leading to varying heat accumulation during current sweeps in opposite directions. Given the temperature sensitivity of superconductors, it is crucial to exclude the potential impact of Joule heating on SDE. Figure 2e displays the *V-I* curves of positive (red solid line) and negative (blue dashed line) current sweep directions. Both curves exhibit a hysteretic *V-I* relationship. The overlapping nature suggests that the sweep direction does not influence the asymmetric *V-I* characteristics, thereby affirming the inherent presence of the field-free SDE in the FTS junction.

Taking advantage of the SDE, a half-wave rectification is conducted at 2 K without external magnetic fields (Figure 2f). A series of square-wave current excitations are applied, with an amplitude of 630 μA, frequency of 0.05 Hz, and rise time of 0.1 ms. Simultaneously, the voltage variation across the junction is recorded in the lower panel. The device maintains a superconducting state under negative current excitation but transitions to a resistive state under positive current excitation. More rectification results can be found in Figure S8 of the Supporting Information.

In previous works aimed at achieving field-free SDE in superconductor junctions, a common approach involved a sandwich-like structure. In this setup, an asymmetric barrier interlayer was consistently intercalated in between two superconductors to facilitate the phenomenon. To explore the underlying mechanisms of the field-free SDE in this FTS junction, Figure 2g displays the cross-sectional STEM image of the junction. The zoom-in focuses on the single FTS nanosheet (Labeled by circled 1) and the overlapping region (Labeled by circled 2), respectively. An oxide layer is identified between the top and bottom FTS layers, as shown in the elemental mapping in the right panel of Figure 2g. This oxide layer exclusively manifests on the upper surface of the FTS nanosheet owing to the fabrication process, while the lower surface of each FTS nanosheet remains intact. This oxide layer exhibits a gradient oxygen concentration, with high oxygen concentration at the upper surface, then gradually diminishes to zero deep into the FTS nanosheet, thus forming an asymmetric oxide barrier.

Based on this asymmetric barrier, we construct the FTS junction model as illustrated in Figure 2h. In the extreme scenario, the asymmetric barrier establishes an insulating-like contact with the top FTS layer (tunneling barrier $\Delta_1$), while forming a metallic contact with the bottom FTS layer (tunneling barrier $\Delta_2$). Different contact methods are employed in Figure S6 of the Supporting Information to elucidate the distinct contact barriers at varying oxygen concentrations in FTS nanosheets. Due to its metallic properties, $\Delta_2$ induces superconducting pairing into the oxide barrier layer from the bottom FTS side, generating a larger proximity region. Taking the interfacial electric field-induced Rashba spin-splitting into consideration, the application of an external voltage modulates the amplitude of the Rashba coefficient, thereby enhancing (or decreasing) pairing under positive (or negative) voltages[34]. Consequently, when the current is applied in the direction that enlarges the proximity region, a higher $I_c$ is achieved, leading to a field-free SDE[34]. More controlled experiments, with and without the oxide barrier, are shown in Figure S5 of the Supporting Information, demonstrating the ubiquity of the field-free diode effect induced by this asymmetric barrier.

The field-free characteristic has been reported in superconducting heterostructures with asymmetric barrier[22] or internal magnetism[35,36]. The former is characterized by a consistent bias in Δ$I_c$, whereas the latter's Δ$I_c$ depends on the history of the applied external magnetic fields. In the FTS junction, Δ$I_c$ remains unchanged after cyclic zero-field cooling (Figure S9, Supporting Information), and after field cooling under positive and negative magnetic fields as large as 8 T (Figure S11, Supporting Information), indicating that internal magnetism does not contribute to the field-free diode effect. Additionally, no magnetoresistance hysteresis loop is observed in FTS nanosheets either below or above the superconducting transition (Figure S10, Supporting Information), further suggesting the absence of internal magnetism in these nanosheets.

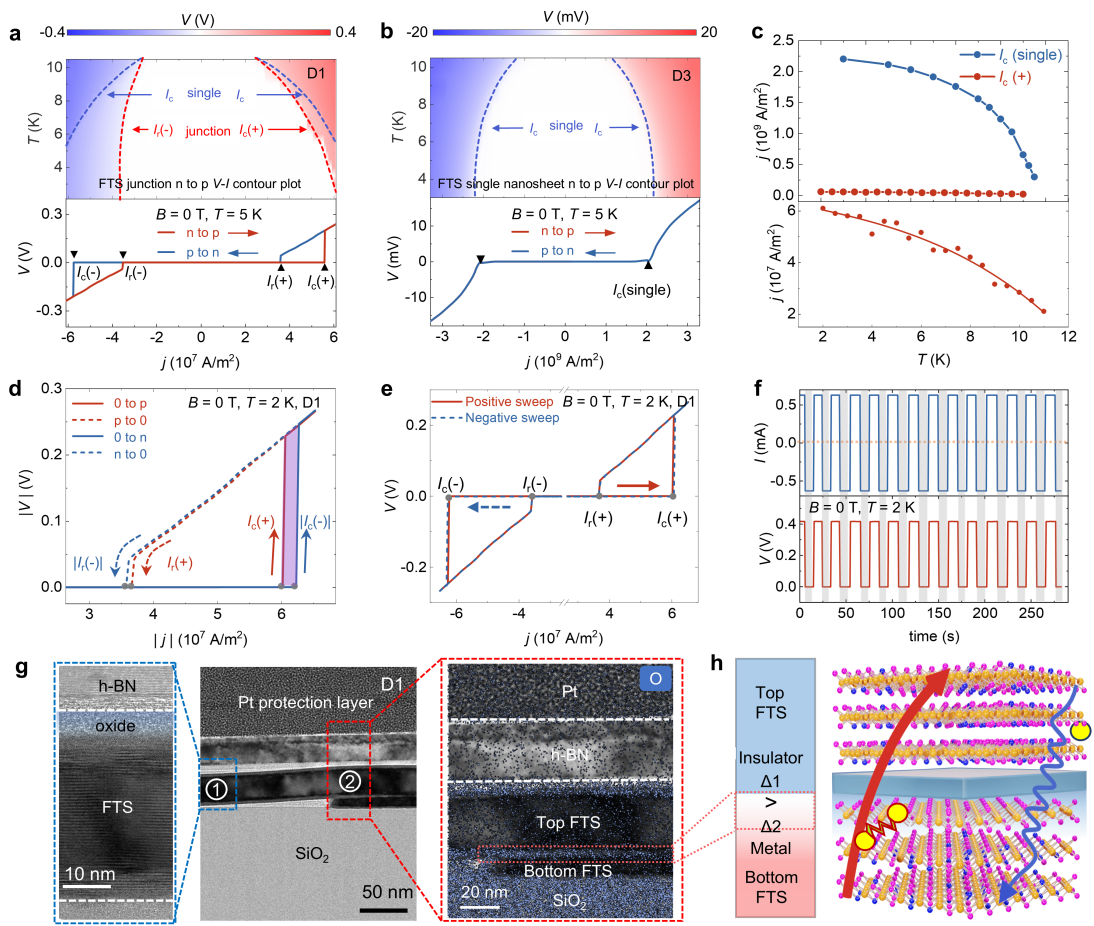

Figure 2. Superconducting diode effect (SDE) of the FTS junction. (a) Upper panel: Voltage contour plot of the FTS junction in the current and temperature plane. Lower panel: *V-I* curve of the junction by sweeping the current in opposite directions. (b) Upper panel: Voltage contour plot of the upper 30 nm FTS nanosheet in the current and temperature plane. Lower panel: *V-I* curve of the single sheet by sweeping the current in opposite directions. (c) Upper panel: Temperature-dependent critical current of the FTS junction and single nanosheet. Lower panel: Zoom in on the critical current of the FTS junction. (d) The absolute value of a *V-I* curve by sweeping the current in the positive direction. (e) *V-I* curves by sweeping the current in a loop in positive (red) and negative (blue) directions. (f) Half-wave rectification of FTS junction. (g) Cross-sectional STEM image of the FTS junction, the left side zooms in on the single sheet region, the right side zooms in on the junction region with oxygen elemental mapping. (c) Schematic of the field-free SDE in FTS junction with an asymmetric barrier.

The absence of inversion symmetry in the FTS junction holds the potential to give rise to magnetic field control of SDE and eMCA. The SDE emerges in the low-temperature regime, where the supercurrent depends on the kinetic energy of Cooper pairs, which reaches its maximum at the lowest temperatures[18]. Meanwhile, eMCA is enhanced at temperatures around $T_c$[9,13], arising from fluctuations in the superconducting order parameters[4]. Despite their distinct temperature dependencies and mechanisms, these two phenomena share similar symmetry requirements. Hence, in most cases, they coexist[14,22], and both respond to external magnetic fields[15,18].

Figure 3a schematically illustrates the nonreciprocal electrical transport measurement setup under magnetic fields. AC current is applied at two ends of the device, while voltages of different harmonic orders across the FTS junction are measured. An external magnetic field is applied and rotates in the yz plane. The misalignment of the lattices between the two FTS nanosheets breaks the inversion symmetry, resulting in spin-splitting, with different Fermi momenta for electrons with opposite spins, as depicted in Figure 3b. An external magnetic field further enlarges the energy dispersion, thereby controlling the magnitude of nonreciprocal responses, manifested in both SDE and eMCA[15,37]. During the superconducting transition, this directional resistance can be described as[2]

$$R(B, I) = R_0(1 + \gamma BI). \tag{1}$$

Harmonic measurement was performed to reveal the role of FTS junction in nonreciprocal electrical transport under magnetic fields of various directions. Under an AC current of $I = \sqrt{2}I_0 \sin\omega t$, the voltage can be expressed as[3,13]

$$V^\omega = \sqrt{2}R_0 I_0 \sin\omega t + R_0 \gamma B I_0^2 \left\{1 + \sin(2\omega t - \frac{\pi}{2})\right\}, \tag{2}$$

where $\gamma$ is the nonreciprocal coefficient, which is an intrinsic parameter of the inversion symmetry-breaking system. By probing the first and second harmonic resistances, we obtain

$$R^\omega = R_0, R^{2\omega} = \frac{1}{\sqrt{2}} R_0 \gamma B I_0. \tag{3}$$

The $R^{2\omega}$ term is correlated with the magnitude of $\gamma$, which is an intrinsic parameter of the system. $R^{2\omega}$ can be modulated by current and magnetic fields, thereby reflecting the magnitude of eMCA.

The harmonic measurements were conducted at 12.5 K, slightly above $T_c$ where the nonreciprocal paraconductivity is enhanced. The upper panel of Figure 3c,d depict the contour plot of $R^\omega$ in the magnetic field strength and direction plane for the FTS junction and single nanosheet, respectively. The lower panel of each figure shows one magnetic field-dependent $R^\omega$ curve to illustrate the superconducting transition under magnetic fields. An increase in the out-of-plane (OOP) magnetic field component tends to drive the device back to the normal state. The two curves exhibit similar shapes, suggesting that the formation of the junction does not significantly influence the behaviour of $R^\omega$.

The upper panel of Figure 3e displays the contour plot of $R^{2\omega}$ in the magnetic field strength and direction plane for the FTS junction, while the lower panel shows one magnetic field-dependent $R^{2\omega}$ curve. The $R^{2\omega}$–$B$ curve exhibits an antisymmetric behaviour, reversing sign under magnetic fields of opposite directions, with the amplitude gradually increasing as the magnetic field tilts towards the OOP direction. In contrast, Figure 3f displays the contour plot of $R^{2\omega}$ for the FTS single nanosheet, while the lower panel specifies one of the magnetic field-dependent $R^{2\omega}$ curves. No $R^{2\omega}$ response is observed for magnetic fields in all directions. Therefore, we believe that the formation of the FTS junction plays a vital role in the realization of eMCA.

This magnetic field dependence applies to both SDE at low temperatures and the nonreciprocal paraconductivity around $T_c$[15]. To reveal this magnetic field dependence under various directions, Figure 3g displays the angle-dependent $\gamma$ (blue dots) and $\Delta I_c$ (red dots), where $\gamma$ is calculated as $\gamma = \frac{\sqrt{2}R^{2\omega}}{R^\omega BI}$. Both $\gamma$ and $\Delta I_c$ respond to external magnetic fields, exhibiting similar trends. The absolute values are maximized under the magnetic field around the OOP direction, while minimized under an IP direction, with a period of $2\pi$.

In this FTS junction, the eMCA and field-dependent SDE are believed to be influenced by the rotational alignment between the two FTS nanosheets. A greater misalignment between the nanosheets is thought to enhance the IP polar vector, leading to a more pronounced magnetic field dependency. Conversely, these phenomena are suppressed in devices where the two FTS nanosheets are aligned parallelly. (Figure S15, Supporting Information) In Figure 3g, the maxima of both $\gamma$ and $\Delta I_c$ emerge under a tilted magnetic field ($\theta = 80°$), deviating from the OOP direction. Conventionally, nonreciprocity is expected to be maximized when the magnetic field is orthogonal to both the polar vector and the current direction. Similar misalignments have also been observed in other junction devices (Figure S13, Supporting Information). This slight misalignment could be attributed to the overlapping configuration of the FTS junction, where the current crossing the junction consists of both IP and OOP components. (Figure S14, Supporting Information)

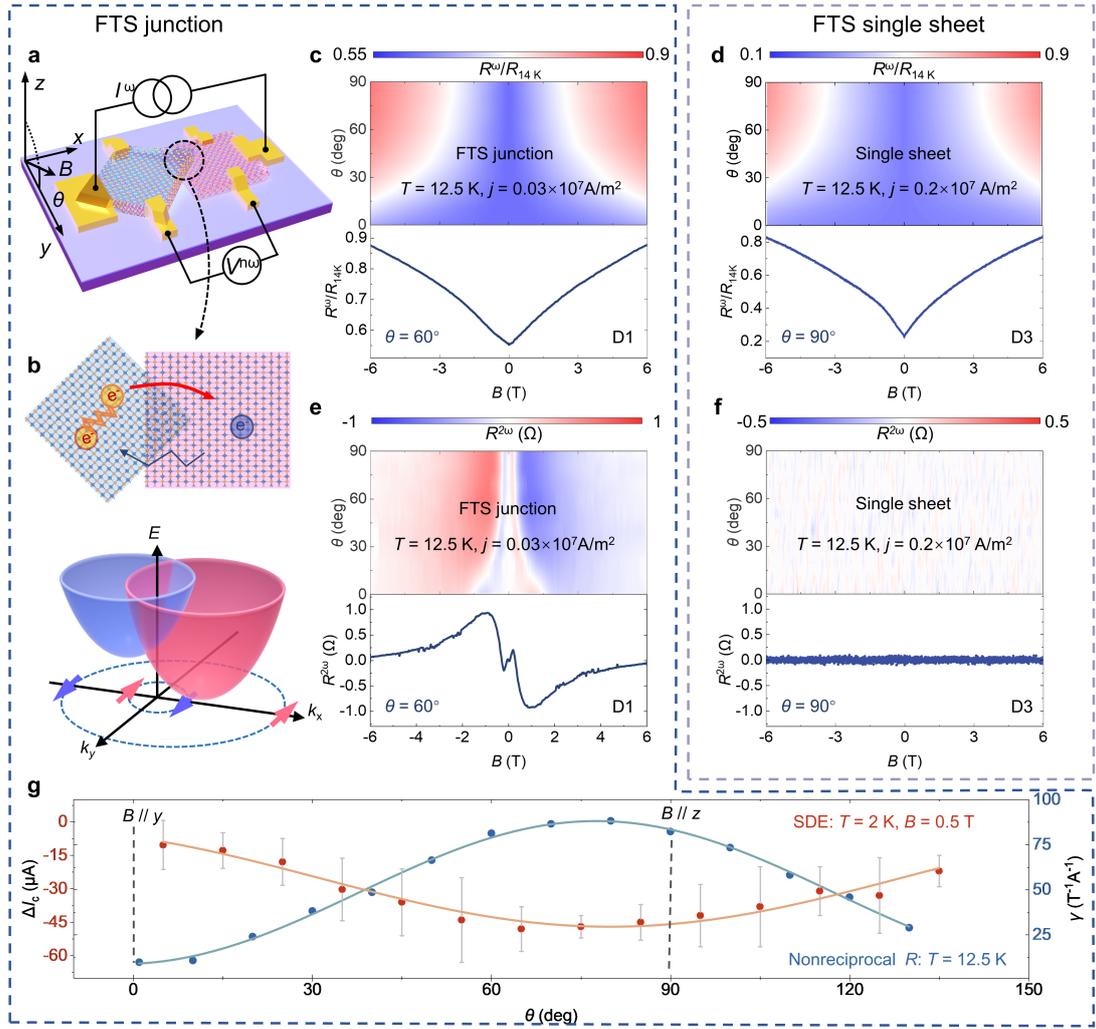

Figure 3. Magnetic field direction dependence of nonreciprocal responses in FTS junction and single nanosheet. (a) Schematic of the harmonic transport measurement under magnetic fields of various directions. (b) Schematic illustration of the structural inversion symmetry breaking of FTS junction and the band dispersion. (c) Upper panel: Contour plot of $R^\omega$ for the FTS junction in the plane of magnetic field strength and direction. Lower panel: Magnetic field-dependent normalized $R^\omega$ of the FTS junction at $\theta = 60°$. (d) Upper panel: Contour plot of $R^\omega$ for the FTS single nanosheet in the plane of magnetic field strength and direction. Lower panel: Magnetic field-dependent normalized $R^\omega$ of the FTS single nanosheet at $\theta = 90°$. (e) Upper panel: Contour plot of $R^{2\omega}$ for the FTS junction. Lower panel: Magnetic field-dependent normalized $R^{2\omega}$ of the FTS junction at $\theta = 60°$. (f) Upper panel: Contour plot of $R^{2\omega}$ for the FTS single nanosheet. Lower panel: Magnetic field-dependent normalized $R^{2\omega}$ of the FTS single nanosheet at $\theta = 90°$. (g) Angle-dependent $\gamma$ at $T = 12.5$ K and $\Delta I_c$ at $T = 2$ K.

The eMCA represents a directional resistance that depends on both current and magnetic field. Particularly, in noncentrosymmetric superconductors, this phenomenon is closely linked with current density and temperature, as they impact the superconducting vortex motion through asymmetric pinning potential[5,12], and the paraconductivity arising from superconductivity fluctuations[13]. Figure 4a presents the current-dependent $R^{2\omega}$ in a wide range of current densities. In the low-current region (purple dashed line), $R^{2\omega}$ slowly increases with current density, attributed to the vortex motion across asymmetric pinning potential across the junction. Higher current increases the driving force, leading to higher $R^{2\omega}$. Subsequently, $R^{2\omega}$ rapidly jumps to nearly ten times higher with a narrow increment in current density (orange dashed line), which is attributed to paraconductivity arising from superconductivity fluctuations at high current densities. A continued increase in current density ultimately results in the suppression of $R^{2\omega}$ due to the destruction of superconductivity (red dashed line), $R^{2\omega}$ is expected to drop to nearly zero above the critical current (red star). Figure 4b focuses on the current-dependent $R^{2\omega}$ in the low current density region, where $R^{2\omega}$ scales linearly with the current density. To demonstrate that eMCA is correlated to the formation of FTS junction, Figure 4c shows the current-dependent $R^{2\omega}$ for the FTS single nanosheet, no $R^{2\omega}$ is observed for all current densities (Figure S18, Supporting Information), but only random noises. This further confirms the efficacy of the formation of the FTS junction in manipulating the spatial and electronic structures.

Recognizing the significant role of the FTS junction in inducing eMCA, we further explore the influence of temperature on this phenomenon. Magnetic field dependence of $R^{\omega}$ and $R^{2\omega}$ at various temperatures below and above the superconducting transition are obtained simultaneously. Figure 4d displays the magnetic field-dependent $R^{2\omega}$ at different temperatures ranging from 7 to 14 K under OOP magnetic fields, while the corresponding $R^{\omega}$ at each temperature is shown in Figure 4e. Two peaks of $R^{2\omega}$ with opposite signs are observed. The first peaks connected by the blue dashed line dominate at low-temperature regime (Peak1), being suppressed to zero above 12.5 K. This peak can be attributed to asymmetric vortex motion from vortex phase fluctuations. The second peaks connected by red dashed lines are negligible at low temperatures, gradually rising and reaching the maximum at 13 K (Peak2). This maximized $R^{2\omega}$ at temperatures slightly above $T_c$ is attributed to paraconductivity from the superconducting order parameter fluctuations.

$\gamma$ values are obtained using the peak value of each $R^{2\omega}$–$B$ curve, along with the corresponding $R^\omega$ and current. $\gamma_1$ represents the flux flow in the vortex phase fluctuation regime, which remains stable at temperatures below $T_c$[38]. With the increase in temperature, the depinning of vortices results in the enhancement of $\gamma_1$, eventually being suppressed with the quench of superconductivity[10]. In contrast, $\gamma_2$ is negligible at low temperatures but is significantly enhanced around $T_c$, a characteristic attributed to paraconductivity[3,13].

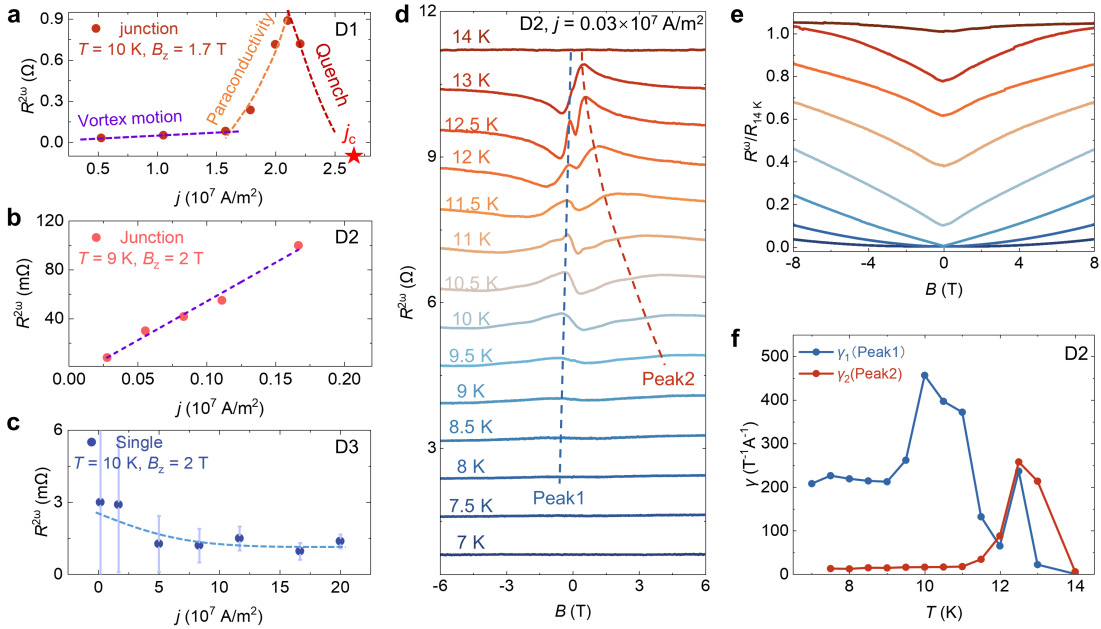

Figure 4. Current and temperature-dependent nonreciprocal responses in FTS junction. (a) Current-dependent $R^{2\omega}$ of the FTS junction over a wide range of current densities. (b) Current-dependent $R^{2\omega}$ of the FTS junction at low current densities. (c) Current-dependent $R^{2\omega}$ of the FTS single nanosheet. (d) Magnetic field-dependent $R^{2\omega}$ at different temperatures ranging from 7 to 14 K under OOP magnetic fields. (e) Corresponding $R^\omega$-$B$ curves at different temperatures. (h) Calculated $\gamma$ as a function of temperature.

## 3. Conclusion

In summary, we demonstrate a field-free SDE accompanied by eMCA in the FTS junction, which is absent in the FTS single nanosheet. The SDE dominates at low-temperature regimes, while the eMCA prevails in the resistive state, exhibiting current and magnetic field dependence at various temperatures. The field-free characteristic is attributed to the bias direction-dependent proximity region through asymmetric tunneling barriers, which arises from an asymmetric oxide barrier layer on the upper surface of the synthesized FTS nanosheets. Meanwhile, the eMCA is believed to arise from spin-splitting induced by the breaking of inversion symmetry through the formation of the FTS junction. Both SDE and eMCA of the FTS junction demonstrate angular dependence on the external magnetic field. This dependence is attributed to the magnetic field-enhanced energy dispersion of spin-splitting[15,16]. This work provides a facile approach for achieving field-free SDE and eMCA in superconductor junctions fabricated from superconducting thin flakes with treated surfaces.

**Experimental Section**

*Sample preparation*

The Tellurium powder (500 mg) and Selenium powders (200 mg) were placed upstream and used as the Te/Se sources. Meanwhile, 10 mg mixed powders with $Fe_2O_3$ and $FeCl_2$ (5:1 by mass) were put in an alumina boat located in the centre of the furnace. The substrates of 280 nm $SiO_2$/Si with polished side faced down were placed downstream of the tube. The carrier gases of 80 sccm Ar and 6 sccm $H_2$ were introduced into the 1-inch quartz tube under ambient pressure. The furnace was heated to 520°C and maintained for 5 mins to synthesize the $FeTe_{1-x}Se_x$ nanosheets. After the growth, the furnace was opened and naturally cooled down to room temperature.

*Device fabrication and transport measurement*

The $FeTe_{1-x}Se_x$ nanosheets on $SiO_2$/Si substrate were transferred via a transfer stage (Perfictlab (Shenzhen) Co., Ltd) to another flake using PDMS stamps. The electrodes were patterned on top and bottom $FeTe_{1-x}Se_x$ flakes using an ultraviolet maskless lithography machine (TuoTuo Technology (Suzhou) Co., Ltd.), followed by electron beam evaporation of Cr/Au (5/85 nm). After lift-off in acetone, the device was capped by h-BN to protect it from degradation. electrical contacts were made by bonding Al wires to the electrodes. The low-temperature transport measurements were conducted in an Oxford TeslatronPT cryostat under a magnetic field of 8 T. The temperature-dependent resistance curves were measured using a Keithley 6221 triggered with a Keithley 2182, at a frequency of 21 Hz. The *V-I* curves were obtained using a Keithley 2400 source meter to apply a DC current and a Keithley 2000 multimeter to measure the voltage. The $R^{\omega}$ and $R^{2\omega}$ were measured simultaneously using a Keithley 6221 source meter and two SR830 lock-in amplifiers with a 17.7 Hz sine wave AC current. Based on the principle, we symmetrized the $R^{\omega}$ raw data and antisymmetrized the $R^{2\omega}$ raw data.

**Supporting Information**

The Supporting Information provides additional details on the elemental analysis of the synthesized FTS nanosheets. Discussions and experimental validations are included to emphasize that the field-free SDE stems from the asymmetric barrier, rather than internal magnetization. Additional controlled experiments are added to demonstrate that the magnetochiral anisotropy in FTS junctions originates from the rotational stacking of two FTS nanosheets, whereas this effect is absent in individual FTS nanosheets.

**Acknowledgements**

X.R.W. acknowledges support from the Academic Research Fund Tier 2 (Grant No. MOE-T2EP50210-0006 and MOE-T2EP50220-0016) and Tier 3 (Grant No. MOE2018-T3-1-002) from Singapore Ministry of Education, and Agency for Science, Technology and Research (A*STAR) under its AME IRG grant (Project No. A20E5c0094). This research is also supported by the Singapore Ministry of Education (MOE) Academic Research Fund Tier 3 grant (MOE-MOET32023-0003) "Quantum Geometric Advantage". Z.L. acknowledges the support from National Research Foundation, Singapore, under its Competitive Research Programme (CRP) (NRF-CRP22-2019-0007 and NRF-CRP22-2019-0004), under its NRF-ISF joint research program (NRF2020-NRF-ISF004-3520). Q.W. acknowledges the support from National Natural Science Foundation of China (Grant No. 52271161). We thank the discussion and support from Prof. Peng Song, and Zhuo Duan.

Table of contents

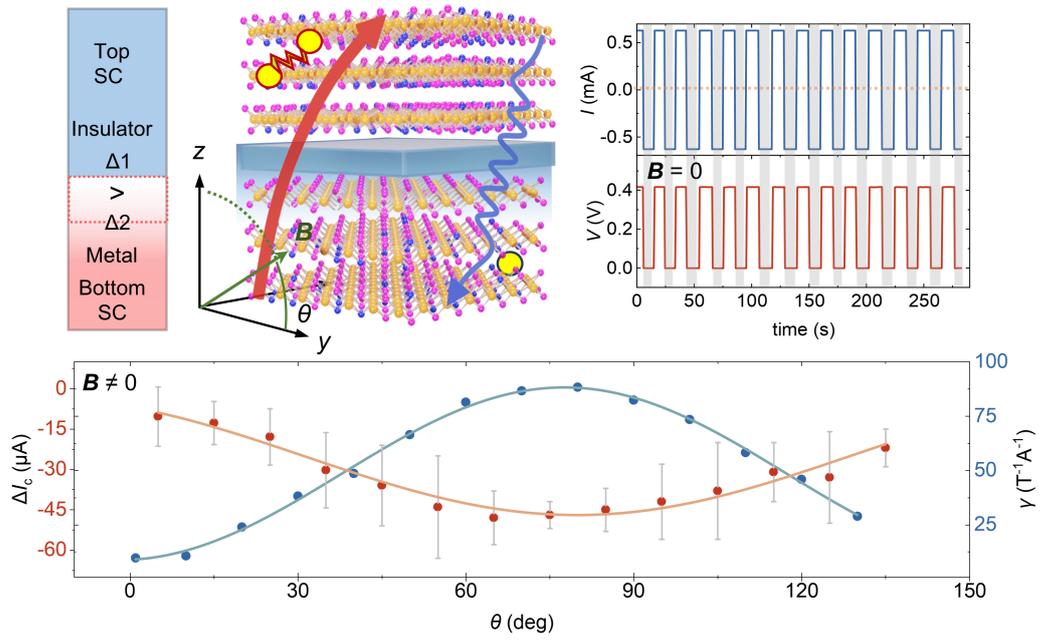